\documentclass[conference]{IEEEtran}
\IEEEoverridecommandlockouts
\usepackage{cite}
\usepackage{amsmath,amssymb,amsfonts}
\usepackage{algorithmic}
\usepackage{stfloats}
\usepackage{graphicx}
\usepackage{textcomp}
\usepackage{xcolor}

\newdimen\extramargin
\graphicspath{{Figures/workflow/}{Figures/CI/}{Figures/hil/}{Figures/levelOfCompliance/}{Figures/objSelectoin/}{Figures/requirements/}}

\def\BibTeX{{\rm B\kern-.05em{\sc i\kern-.025em b}\kern-.08em
    T\kern-.1667em\lower.7ex\hbox{E}\kern-.125emX}}
\begin{document}
\title{A Lean and Highly-automated Model-Based Software Development Process Based on DO-178C/DO-331
}

\author{\IEEEauthorblockN{Konstantin Dmitriev, Shanza Ali Zafar, Kevin Schmiechen, Yi Lai, Micheal Saleab, Pranav Nagarajan, \\Daniel Dollinger, Markus Hochstrasser and Florian Holzapfel}
	
	\IEEEauthorblockA{\textit{Technical University of Munich}\\ Garching, Germany \\
		\{konstantin.dmitriev,
		shanza.zafar,
		kevin.schmiechen,
		yi.lai,
		micheal.saleab,
		pranav.nagarajan,\\
		daniel.dollinger,
		markus.hochstrasser,
		florian.holzapfel\}@tum.de} 
	
	\\
	
	\IEEEauthorblockN{Stephan Myschik}
	\IEEEauthorblockA{\textit{Universitaet der Bundeswehr Muenchen}\\ Neubiberg, Germany \\
		stephan.myschik@unibw.de}
}

\maketitle
\IEEEpeerreviewmaketitle

\begin{abstract}
 The emergence of a global market for urban air mobility and unmanned aerial systems has attracted many startups across the world. These organizations have little training or experience in the traditional processes used in civil aviation for the development of software and electronic hardware. They are also constrained in the resources they can allocate for dedicated teams of professionals to follow these standardized processes. To fill this gap, this paper presents a custom workflow based on a subset of objectives derived from the foundational standards for safety critical software DO-178C/DO-331. The selection of objectives from the standards is based on the importance, degree of automation, and reusability of specific objectives. This custom workflow is intended to establish a lean and highly automated development life cycle resulting in higher quality software with better maintainability characteristics for research and prototype aircraft. It can also be proposed as means of compliance for software of certain applications such as unmanned aircraft systems, urban air mobility and general aviation. By producing the essential set of development and verification artifacts, the custom workflow also provides a scalable basis for potential future certification in compliance with DO-178C/DO-331. The custom workflow is demonstrated in a case study of an Autopilot Manual Disconnection System.

\end{abstract}


\section{Introduction}
Software is an essential part of modern air vehicles and it usually implements complex safety-critical functions. To ensure safety of airborne software, the aviation community uses the standard DO-178C \cite{RTCA.178} and its supplements (e.g. DO-331 \cite{RTCA.331} addressing model-based aspects) as a primary guidance for airborne software development. They describe in detail "objectives for software life cycle processes, activities that provide a means for satisfying those objectives, descriptions of the evidence in the form of software life cycle data [...] [and] variations in the objectives [...] by software level [and] additional considerations (for example, previously developed software)" \cite[p.~1]{RTCA.178}. However, achieving compliance with DO-178C significantly increases cost and time of development. The growth  of project expenses for DO-178C level C software can exceed 150\% compared to the development of typical high-quality consumer software per Capability Maturity Model
Integration (CMMI) Level 2 or 3 \cite{thomas2009certCost}. While this expense may be acceptable to some extent for large transport aircraft (given the perception of risk in this category), it is a major hindrance to the advancement of technology -- especially avionics -- in normal category airplanes and rotorcraft. The availability of cost-effective qualified avionics solutions for these categories could prevent several general aviation (GA) accidents, such as controlled flight into terrain in poor weather or loss of control. This challenge has received increased visibility in recent years, owing to the growth of rapidly emerging and innovative markets such as urban air mobility (UAM) and unmanned aircraft systems (UAS).

Manufacturers and component suppliers developing avionics solutions for GA, UAS/UAM are cooperating with certification authorities to develop alternative means of compliance for software assurance other than DO-178C. The recently created EUROCAE and RTCA joint working group WG-117 is targeting to develop a software standard that will be specific to lower risk UAS and GA applications \cite{EUROCAEwg117}. ASTM International has released standards on the following topics intended to streamline avionics certification: system-level verification to assure safety of the intended function (ASTM F3153 \cite{ASTM.3153}), run-time assurance of complex functions (ASTM F3269 \cite{ASTM.3269}), dependability of software used in UAS (ASTM F3201\cite{ASTM.F3201}). These standardization development efforts target applicants with little or no experience with certification, especially in the aerospace industry. At the same time, it is also expected that software certification for high-risk operations in UAS/UAM will still require conformity to DO-178C processes.

Many developers also use rapid prototyping methods during their research, development and flight testing phases. With constrained resources, a lack of knowledge and a still developing certification environment for their products, these companies are looking for processes as agile as their development cycles. However, certain DO-178C objectives make it difficult to reuse software previously developed out of DO-178C context. For example, in case of poor traceability between requirements of all levels down to source code, it may be easier to redo the development from scratch, rather than re-engineer trace data, improve requirements and rework source code accordingly. This aspect also contributes towards the lean though scalable development life cycle for prototype and experimental airborne software. For such categories of software products, we deem it beneficial to establish a lean but scalable software development life cycle, which helps to achieve a high level of software quality and provides a basis for potential reuse of experimental software for further certification in accordance with DO-178C. 

Use of advanced development techniques like model-based design \cite{Marcil.2012.MBD} or agile process models \cite{coe2013model} along with automation of development and verification activities with qualified tools \cite{hochstrasser2018process} can help to reduce the development effort while achieving full compliance with DO-178C and DO-331. Below, we discuss the lean model-based software development life cycle (\textit{custom workflow}), which utilizes these advance techniques along with a further reduction of the development effort by introducing the \textit{custom subset} of DO-178C and DO-331 objectives. We selected the objectives for the \textit{custom subset} based on criteria of importance, automation and reuse with the following goals: 
\begin{enumerate}
	\item Reduce non-critical process activities to minimize development efforts while maintaining a software quality level similar to DO-178C level C.
	\item Enable the completion of software development to achieve full compliance with DO-178C per level C or higher in the future without wide rework of the data produced using the custom workflow. 
\end{enumerate}

This paper is structured in the following sections. Section \ref{sec:objSel} describes the methodology for the selection of the \textit{custom subset} of objectives. In section \ref{sec:wrkOvr}, a general overview of the workflow to satisfy the selected objective using \textit{MathWorks} and own tools is given. This is followed by an explanation of in-house processes and tools to facilitate the \textit{custom workflow} in sections \ref{sec:requirementsManagement} to \ref{sec:HIL}. Section \ref{sec:lvlComp} describes the achieved level of compliance to DO-178C and DO-331 using the \textit{custom workflow}. A case study is performed by developing an \textit{Autopilot Manual Disconnection System} (AMDS) using the proposed workflow and is described in section \ref{sec:caseStudy}. Section \ref{sec:futureWrk} discusses the planned enhancement for the \textit{custom workflow}. At the end, conclusions are given in section \ref{sec:conclusions}.

\section{Objectives Selection Concept}\label{sec:objSel}
We considered the following criteria to select the DO-178C and DO-331 objectives for the \textit{custom subset}:
\begin{itemize}
	\item Importance of the objective
	\item Level of automation for objective implementation
	\item Level of potential reuse of objective data (artifacts)
\end{itemize}

\textbf{Importance of the objective} was evaluated based on following factors:
\begin{enumerate}
	\item In DO-178C, the objectives that need to be satisfied during software development depend on software Levels A to E resulting from the safety assessment process. Level A corresponds to the most critical \textit{Catastrophic} failure condition category and requires most objectives to be satisfied with maximum rigor while Level E corresponds to \textit{No Safety Effect} and requires no objectives to be satisfied. Objectives applicable to all software levels from A to D are considered more important than the objective assigned to only higher levels (i.e. level C or higher). Examples include the definition of high-level software requirements and testing of executable object code for compliance with high-level requirements.
	\item Relevance of the objectives to other aerospace and non-aerospace software development standards (ASTM F3201 \cite{ASTM.F3201}, ISO 26262-6\cite{ISO.26262-6}, ISO 26262-8\cite{ISO.26262-8}, EN 50128\cite{EN.50128} have been considered). For example, traceability down to source code is not required by DO-178C/DO-331 for Level D, but EN 50128 and ISO 26262-8 recommend the complete traceability down to source code.
\end{enumerate}

\textbf{Level of automation} was another key factor for selecting the objectives for the \textit{custom subset}. If the objective can be automated using a tool, it can help to improve overall software quality without significant increase of manual efforts. Examples include:
\begin{itemize}
\item Requirements verification objectives (tables MB.A-3 and MB.A-4 of DO-331) which can be substantially automated for model-based software (MBS) components leveraging model simulation. Test cases and procedures developed for model simulation can be further seamlessly reused for executable code testing.
\item Code verification objectives (table A-5/MB.A-5 of DO-178C/DO-331). These objectives are not applicable to lower software levels (Level D and below) and hence are not assigned with high importance as per the criteria above. However, they can be efficiently automated leveraging \textit{Simulink Code Inspector}\footnote{https://www.mathworks.com/products/simulink-code-inspector.html}\cite{conrad2012slci}, \textit{Polyspace Bug Finder}\footnote{https://www.mathworks.com/products/polyspace-bug-finder} and \textit{Polyspace Code Prover}\footnote{https://www.mathworks.com/products/polyspace-code-prover} tools with very little additional manual effort.
\item Objectives for code structural coverage (table MB.A-7 of DO-331), which can be automated for MBS code, reusing the test cases for model simulation and applying the \textit{Simulink Coverage tool}\footnote{https://www.mathworks.com/products/simulink-coverage}.
\end{itemize}

\textbf{Level of potential reuse} was considered as another factor due to the increased value of the artifacts, which can be leveraged without major modification across multiple projects. The value of such artifacts and corresponding objectives versus efforts spent for their development is multiplied by reuse and therefore such objectives are considered more relevant to the \textit{custom subset}. For example, certain parts of project development and verification plans and standards can be written in a generic way to be reusable for other development projects being implemented in a similar environment. 

\begin{figure}[hbt!]
	\centering
	\includegraphics[width=\columnwidth]{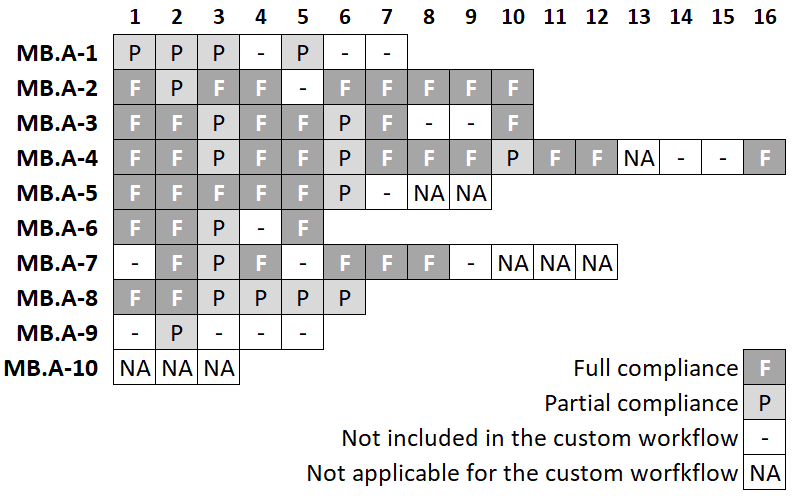}
	\caption{DO-331 Compliance Summary for Model-based Software}
	\label{fig:do331complianceSummary}
\end{figure}

\begin{figure}[hbt!]
	\centering
	\includegraphics[width=\columnwidth]{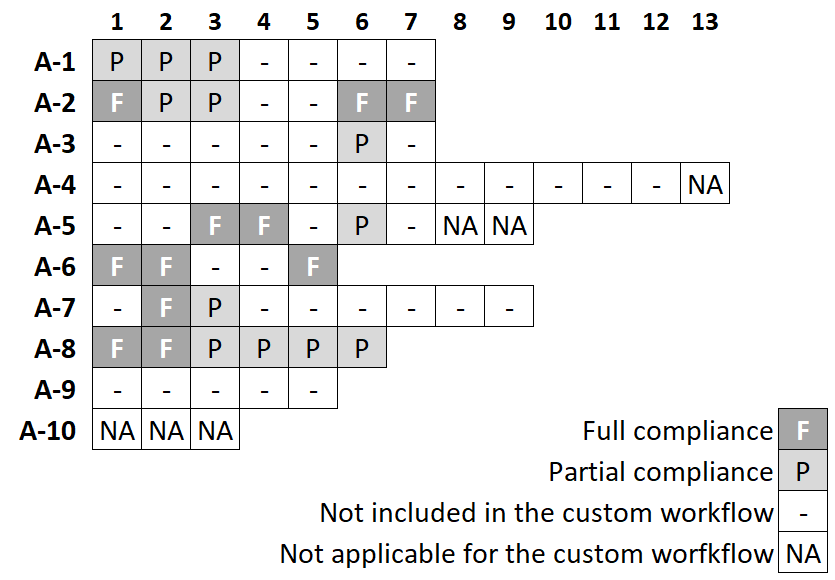}
	\caption{DO-178C Compliance Summary for Low-level Software}
	\label{fig:do178complianceSummary}
\end{figure}

We evaluated DO-178C and DO-331 objectives and assigned each objective with a high, medium and low level of applicability for each selection criteria. Based on the total score qualitatively assessed as a combination of all three criteria levels, we selected DO-331 and DO-178C objectives for the \textit{custom workflow}. This qualitative assessment was based on consensus between the authors. Fig. \ref{fig:do331complianceSummary} and \ref{fig:do178complianceSummary} show the summary of the selected DO-331 objectives applicable to MBS components and DO-178C objectives applicable to manually coded low-level software (LLS) components respectively. Some of the objectives are not applicable to the \textit{custom workflow} and shown as N/A on the figures. For example, certification liaison objectives are not applicable because a formal certification process is not implemented in the \textit{custom workflow}. These figures also indicate the compliance level achieved for each objective in the implemented \textit{custom workflow}. This is further discussed in section \ref{sec:lvlComp}.

\section{Workflow Overview}\label{sec:wrkOvr}
\begin{figure*}[!hbt]
	\centering
	\includegraphics[width=\textwidth]{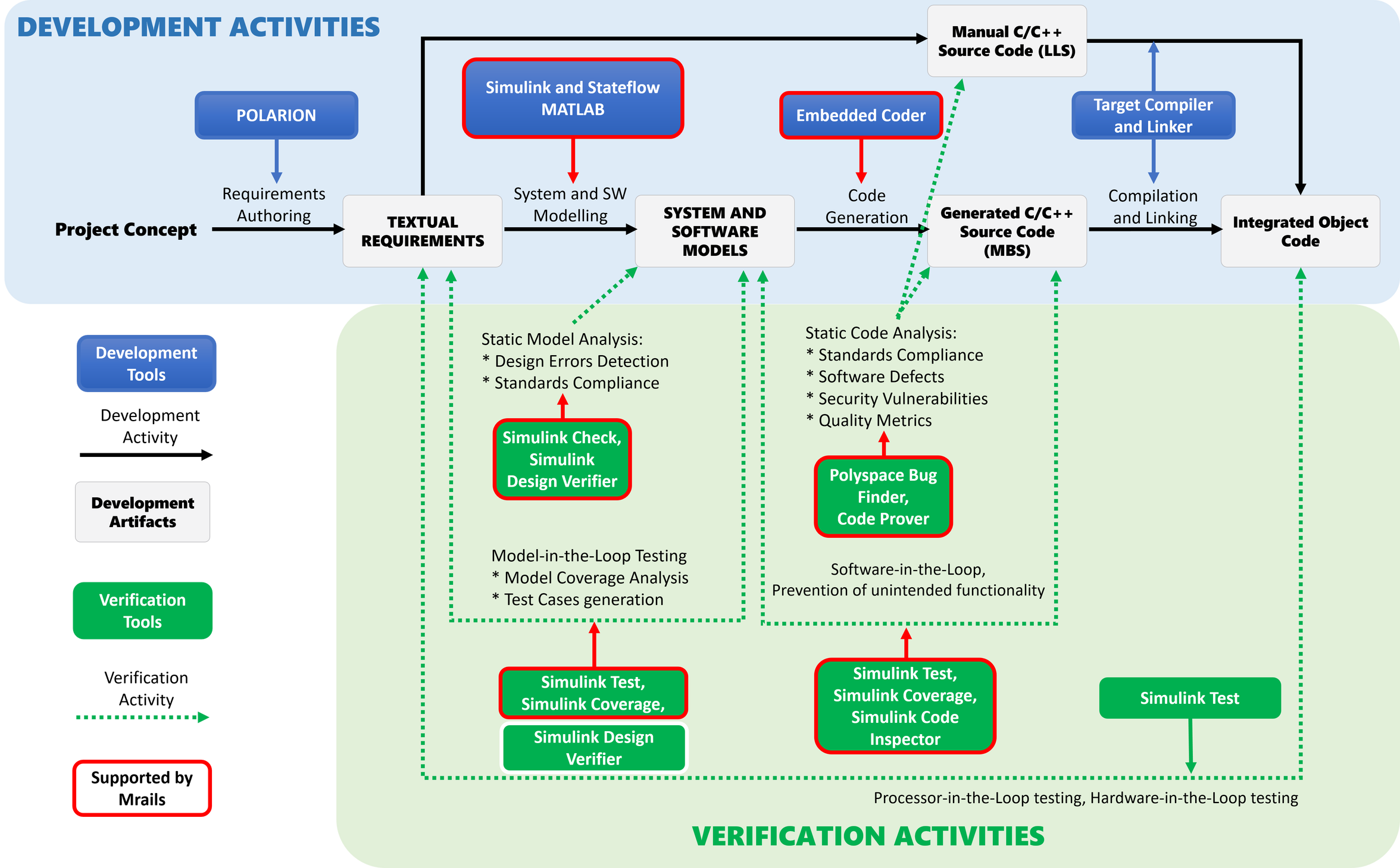}
	\caption{Development and verification workflow using MathWorks based toolchain for safety critical systems}
	\label{fig:WF}
\end{figure*}

The workflow to achieve the \textit{custom subset} of DO-178C/DO-331 objectives discussed in section \ref{sec:objSel} can be structured in these parts:
 
 \begin{itemize}
 	\item Planning activities for applicable objectives corresponding to table A-1/MB.A-1 of DO-178C/DO-331.
 	\item Development activities corresponding to table A-2/MB.A-2 of DO-178C/DO-331. 
 	\item Verification activities corresponding to table A-3/MB.A-3 through A-7/MB.A-7 of DO-178C/DO-331.
 	\item Configuration management and quality assurance activities corresponding to tables A-8/MB.A-8 and A-9/MB.A-9 of DO-178C/DO-331.
\end{itemize} 
We are using  \textit{MathWorks} products for automation of many objectives due to legacy projects and supported \textit{MathWorks}-based solutions but the workflow can also be adapted for other commercial Model-based Development (MBD) toolchains, for instance \textit{Ansys SCADE Suite}\footnote{https://www.ansys.com/products/embedded-software/ansys-scade-suite}. An overview of the development and verification activities for the software development workflow is shown in Fig.~\ref{fig:WF}. Configuration management and quality assurance are applied for all artifacts and not shown on the figure for simplicity.

\textbf{Planning activities} can require significant manual effort but the reusability of artifacts across different projects is high as mentioned in section \ref{sec:objSel}. For our lean \textit{custom workflow}, planning activities include defining the life cycle in form of an interactive user guide detailing the activities to be performed. This guide also includes the description of the tools to be used and how to use them for MBS and LLS components. Another aspect of the planning process are software development standards. For MBS, DO-331 also requires a \textit{Software Model Standard}. As a part of this, \textit{Modeling Guidelines} based on a subset of \textit{MathWorks} modeling guidelines \cite{modelingguidelines} are developed at our institute for \textit{Design Models}. These are models which are used to generate source code directly. Additional guidelines are also added to overwrite or further restrict functionalities based on experience gathered during previous projects (e.g. for testing or to avoid rounding errors and overflows). \textit{Naming Conventions} are also developed at our institute for consistency and compliance with MISRA C \cite{misra} and the \textit{Modeling Guidelines} with a focus on flight dynamics and control. 

\textbf{Development Activities'} entry point are textual requirements authored in \textit{Polarion}. In model-based development, a textual requirement may represent different levels of system and software requirements, depending on the selected scenario of model usage as described in DO-331 MB.1.6.3. The \textit{custom workflow} is implemented  based on the MB Example 5 practice (see DO-331 Table MB.1-1) for MBS components. We capture system requirements in textual format and further  implement them as \textit{Design Models} in \textit{Simulink} for the MBS part (implementing functional capabilities) and refine textual software requirements for the LLS part (scheduling layer, hardware interfaces). \textit{Simulink} models represent low-level requirements and software architecture. Source code is then automatically generated from \textit{Design Models} using \textit{Embedded Coder}. Further details about the modeling environment used for \textit{Design Models} can be found in \cite{hochstrasser2018aspects}.

For the LLS part, source code is manually written based on the refined LLS requirements and intended to be generally reused across projects implemented on the same hardware. Finally, LLS code is integrated with the auto-generated code for MBS code and compiled to get target compatible object code. 

\textbf{Verification activities} can be greatly automated and performed early for MBS using \textit{Design Models}. DO-331 allows simulations as means of compliance for the key objectives in table MB.A-3 (Verification of Outputs of Software Requirement Process) and further objectives in table MB.A-4 (Verification of Outputs of Software Design Process). Example of these objectives include \textit{high-level requirement comply with system requirements, requirements are accurate and consistent, algorithms are correct}). Requirement-based simulation cases can be further fully reused for software-in-the-loop (SIL) and Hardware-in-the-loop (HIL)  testing of the executable object code to satisfy some objectives of table MB.A-6 (Testing of Outputs of Integration Process). Using \textit{Simulink Test}\footnote{https://www.mathworks.com/products/simulink-test.html} for simulation and testing helps achieving objectives in table MB.A-7 (Verification of Verification Process Results) by facilitating structural coverage collection, requirement traceability, and reviewing and managing test results. Reusability of test cases for HIL using \textit{Simulink Test} is further discussed in subsection \ref{sec:HIL}. 

Checking models during development for design errors using \textit{Simulink Design Verifier}\footnote{https://www.mathworks.com/products/simulink-design-verifier} and for conformance to the \textit{Software Model Standard} using \textit{Simulink Check}\footnote{https://www.mathworks.com/products/simulink-check} helps fulfilling other objectives of MB.A-3 and MB.A-4 (Verification of Outputs of Software Design Process). Furthermore, some objectives corresponding to the verification of source code (both manually coded for LLS and auto-generated for MBS) in table A-5/MB.A-5 (Verification of Outputs of Software Coding and Integration Process) are automated using static code analysis techniques for compliance with standards like MISRA C, detecting run-time errors and for proving absence of some run-time error through \textit{Polyspace Bug Finder} and \textit{Polyspace Code Prover}. Additionally for MBS, objectives for source code compliance to low-level requirements and software architecture in table MB.A-5 are accomplished by \textit{Simulink Code Inspector}. 

\textbf{Configuration management activities} are implemented by using \textit{Polarion} for requirements management and \textit{GitLab} for other artifacts. These tools implement essential activities required for the DO-178C/DO-331 \textit{Control Category 2}: identification, traceability and change tracking. More stringent activities required for the DO-178C/DO-331 \textit{Control Category 1} may hinder rapid prototyping and are not included in the \textit{custom workflow}.

\textbf{Quality assurance activities} are implemented using the tool \textit{mrails} described below and are limited to tracking the completion status of the planned activities. For example, the left panel of the \textit{mrails} status overview window presented in Fig.~\ref{fig:WF} shows the status of the planned workflow activities.

The proposed \textit{custom workflow} enables high automation in both development and verification activities owing to MBD, which puts the foundation for a lean workflow based on DO-178C. Few examples of MBD for safety critical systems in literature are \cite{estrada2013best,krizan2014automatic,erkkinen2007safety}. 

In the following subsections, we will discuss the specific solutions employed at the institute which facilitate the implementation of the lean \textit{custom workflow} in addition to the \textit{MathWorks} tools mentioned above.

\subsection{Requirements Management}\label{sec:requirementsManagement}
This section presents the requirement management setup for our \textit{custom workflow}. We used the tool \textit{Polarion}, which is a web-based application life cycle management solution by \textit{Siemens Digital Industry Software}, to manage the textual requirements and their trace data. While writing the requirements, we applied the \textit{FRETISH} language of the tool \textit{FRET} (Formal Requirements Elicitation Tool) as much as possible. It was developed at \textit{NASA Ames Research Center} and was presented in \cite{Giannakopoulou2020}. The tool \textit{SimPol}, which was developed at our institute, is used to create bidirectional traceability between requirements on \textit{Polarion} and \textit{Design Models} and test cases in \textit{Simulink Test}. Further details in the context of requirement derivation, simulation and validation are discussed in \cite{Schmiechen.2019, Meidinger.2019}.

\begin{figure*}[hbt!]
	\centering
	\includegraphics[width=\textwidth]{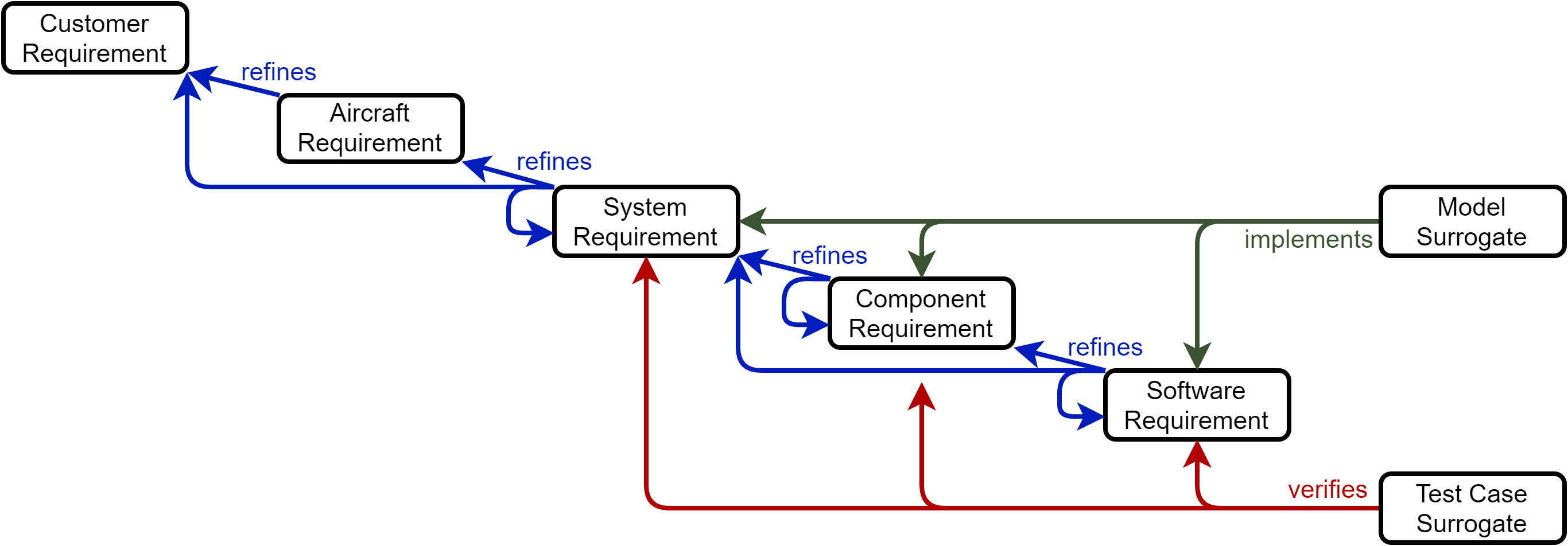}
	\caption{Work Item Types and Link Roles}
	\label{fig:WorkItemDef}
\end{figure*}

\textbf{Artifact Hierarchy} for our \textit{custom workflow} is shown in Fig.~\ref{fig:WorkItemDef}. It includes the work item types (e.g. requirements) but also the link roles. Lower level requirements refine upper level requirements, however, as a simplification, it is also possible that requirements refine those from the next but one level (e.g. from software to system) or that requirements refine other requirements from the same level (e.g. component) because a higher level of detail is necessary within one level. Even though the arrows are only pointing in one direction, the traces are still bidirectional (e.g. forward trace: refines / backward trace: is refined by). Additionally, review checklists are used to hold traceability evidence based on the automatic trace reports described below. The other verification objectives of tables A-3/MB.A-3 and A-4/MB.A-4 of \textit{DO-178C/DO-331}, which require manual review, can be included in the review checklist when increasing the compliance level with respect to the standard.

\textbf{Cross-Platform Artifact Linking} is done with \textit{SimPol} between the requirements on \textit{Polarion} and the \textit{Design Models} and test cases in \textit{MATLAB/Simulink}. The tool creates model and test case surrogate work items as described in Fig.~\ref{fig:WorkItemDef}, which serve as a representation on \textit{Polarion} for the actual models and test cases. Automatic highlighting of links where either side has changed enables an impact analysis of such changes. The available links can also be used to identify the impact of changes to other work items before the changes are even made.

\begin{figure}[hbt!]
	\centering
	\includegraphics[width=\columnwidth]{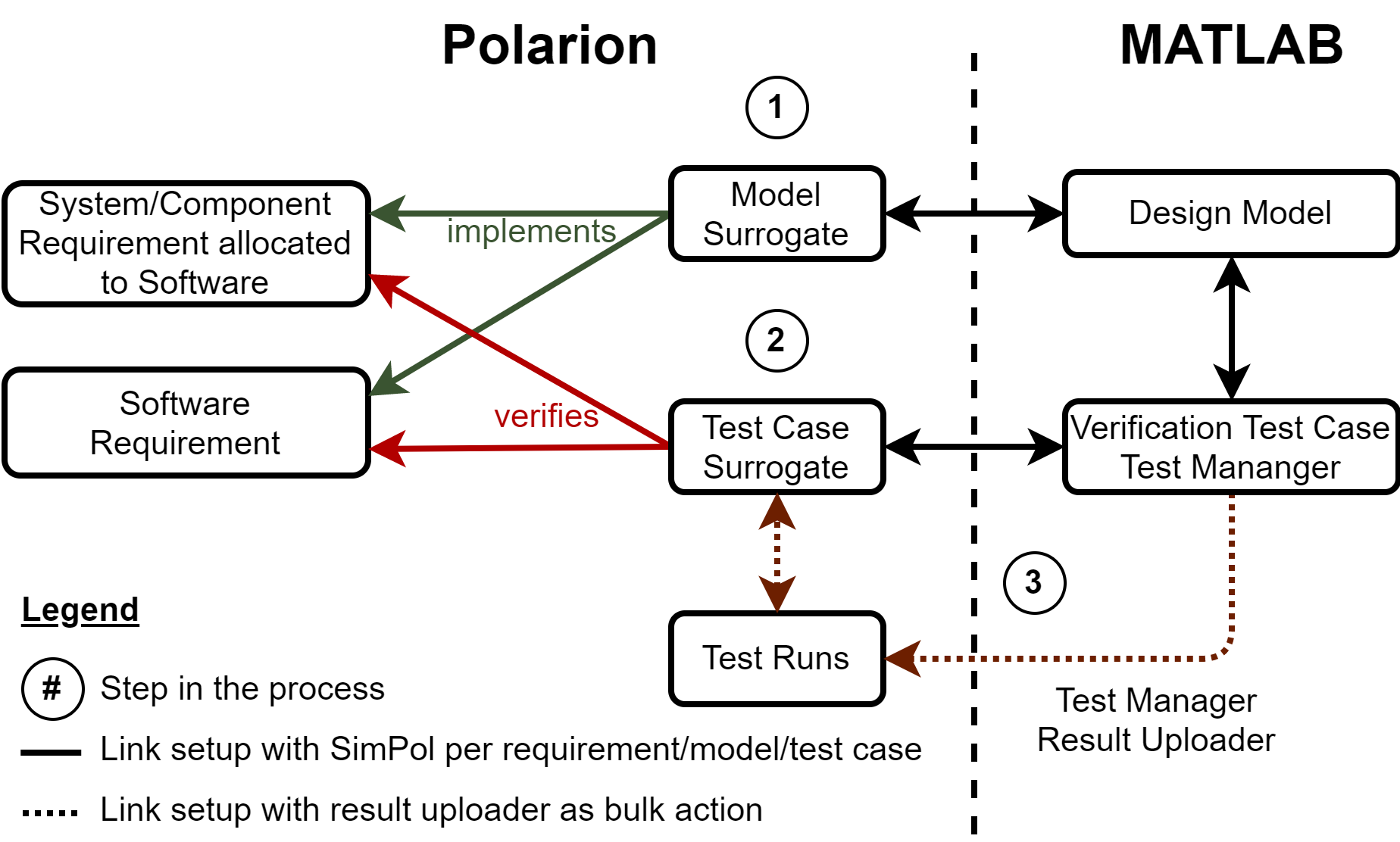}
	\caption{Requirement, Model and Test Case Linking}
	\label{fig:ReqModelTestCaseLinking}
\end{figure}

In accordance with the wording of \textit{DO-331}, the \textit{Design Models} have the link role \textit{implements} to software requirements and to those component and system requirements that are allocated to software. Model-based test cases have the link role \textit{verifies} to requirements on \textit{Polarion}. An overview of the steps to link between \textit{Polarion} and \textit{Simulink Test} is given in Fig.~\ref{fig:ReqModelTestCaseLinking}. In the first two steps, the requirements on \textit{Polarion} are linked with the \textit{Design Models} in \textit{Simulink} and the test cases in the \textit{Simulink Test Manager}. The link from model-based test case to \textit{Design Model} is already established during the creation of the test case and its harness. The third step is to upload the test run records to \textit{Polarion}. A custom routine processes the records from \textit{Simulink Test Manager} into an \textit{XML} file. A custom extension on the \textit{Polarion} server processes this \textit{XML} file, creates the test run records and links them to the test case surrogates.

\textbf{Automatic Trace Reports} can be generated on \textit{Polarion} with a custom widget as soon as the work items are created and linked with each other. These reports correspond to parts of \textit{Trace Data} required by section 11.21 of \textit{DO-178C}. In addition to the actual trace table, the report also consists of a pie-chart with the distribution of the trace status of the work items for a quick overview. Derived requirements, which are identified by a custom field, are one example where it is justified that a requirement does not have an upstream trace.

The custom widget was inspired by a widget for test case coverage that is part of the default \textit{Polarion} installation, however, our widget was developed to be applicable to our requirement traceability and even more use cases. It also supports checking traces in both directions (e.g. also from lower to higher level requirements).

\subsection{Automation of Life cycle Activities}\label{sec:mrails}
A process-oriented build tool (\textit{mrails}) for MBS using the \textit{MathWorks} toolchain was developed at our institute by Hochstrasser \cite{hochstrasser2018process} which greatly reduces the efforts required for certification-ready software development by automating the life cycle activities discussed in section \ref{sec:wrkOvr}. In Fig.~\ref{fig:WF}, activities supported and automated by \textit{mrails} in the overall workflow are highlighted.

For the development of \textit{Design Models}, \textit{mrails} integrates the modeling environment and supports the creation of \textit{'ready-by-construct'} building elements such as reusable models, or \textit{Simulink} bus objects compatible with \textit{Modeling Guidelines}. 

\textit{Mrails} also automates the execution of verification activities, named \textit{jobs}, for \textit{Design Models} and auto-generated source code. Furthermore, it supports the generation of manual reviews lists for verification such as to check compliance with \textit{Modeling Guidelines} which cannot be automated by \textit{Simulink Check}.

 In case of software changes after certification, the question which artifacts needs to be reevaluated often leads to a \textit{'Big Freeze'} problem (i.e. baselining the software)\cite{hochstrasser2019processapplication}. A solution to this is implemented in \textit{mrails}. For all jobs, an artifact dependency graph is generated between the manually modifiable \textit{source files} like requirements, models and the auto-generated \textit{derived artifacts} like reports, test results. \textit{Mrails} uses the dependency graph to perform a change impact analysis. This analysis identifies the affected artifacts that are needed to be regenerated. Running the analysis again only generates the missing or outdated artifacts, which is faster than to brute-force regenerate all artifacts. This feature of \textit{mrails} facilitates the continuous and incremental certification approaches mentioned in \cite{comar2009open}.

\textit{Mrails} provides a web-based interface summarizing the results of jobs along with the status showing the validity of dependent artifacts (i.e. whether the source has changed and the artifacts are outdated). The advantage of this interface is that it can be used outside the \textit{MATLAB} environment and helps in the reviewing process. In \cite{hochstrasser2019processapplication}, an example using \textit{mrails} for flight control development is provided.

\textit{Mrails} interacts well with Continuous Integration (CI) by providing a simple MATLAB command-line interface that can be easily used on a CI server.

\subsection{Continuous Integration}\label{sec:ci}
Continuous Integration (CI) is a common practice nowadays in software development. It provides the benefit of integrating features more frequently, e.g. daily, by automating building and testing as much as possible and detects integration errors in early stages. In our model-based software development process, we use CI to extensively automate the execution of model static analysis, simulation-based testing, unit testing and static analysis of manual source code. With \textit{Mrails}, which was described in section \ref{sec:mrails}, the CI requires only minimum configuration since most of the tasks are as simple as running a \textit{job} in \textit{Mrails}. While many CI tools and services, e.g. \textit{Jenkins}\footnote{https://www.jenkins.io/}, \textit{Travis CI}\footnote{https://travis-ci.com/}, provide similar generic features that are used in our process, we have chosen \textit{Gitlab CI/CD} as CI tool because it is natively supported by the source control tool \textit{GitLab} that was already in use at our institute.

\begin{figure}[hbt!]
	\centering
	\includegraphics[width=\columnwidth]{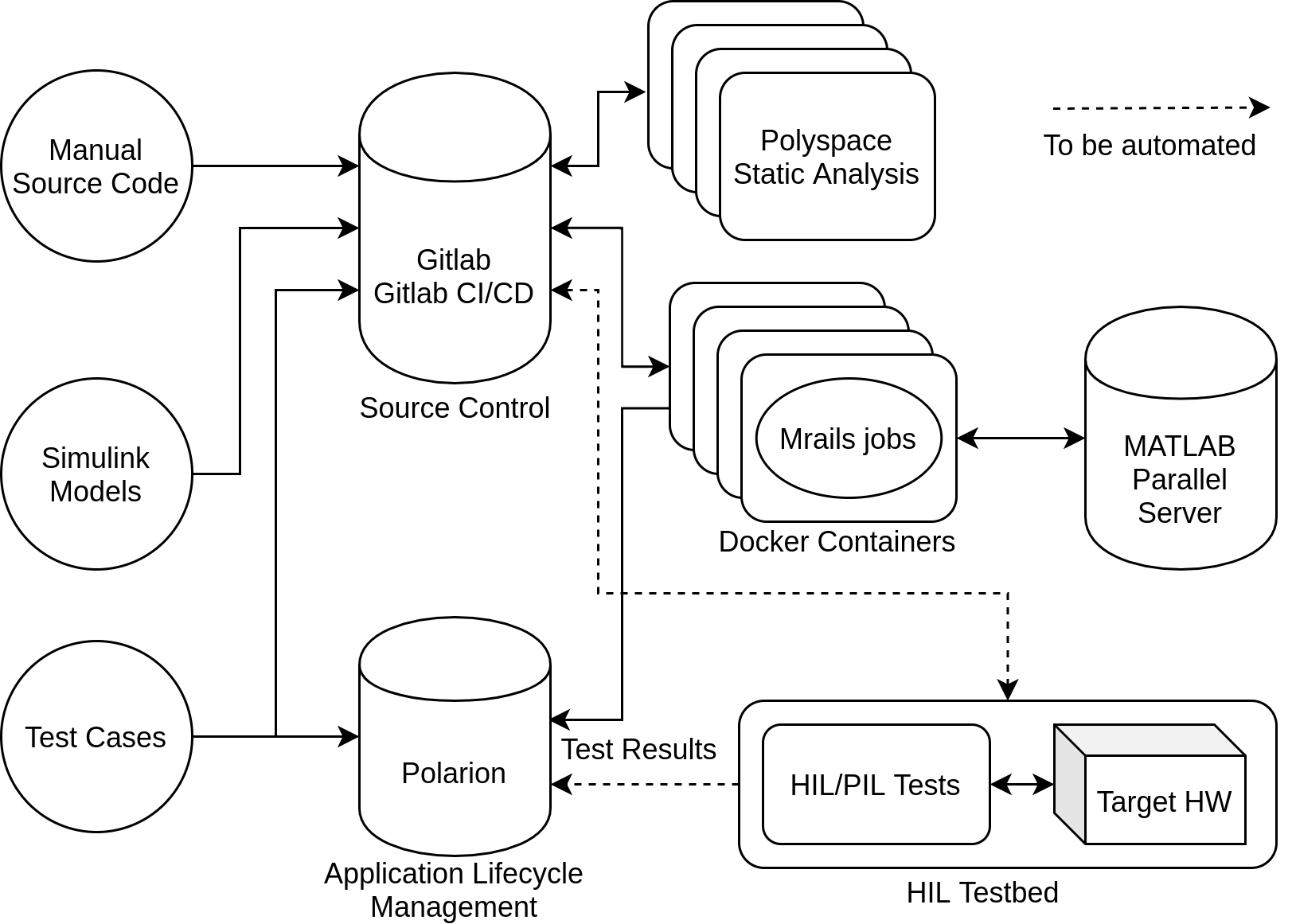}
	\caption{Continuous Integration Overview}
	\label{fig:CI_overview}
\end{figure}

\textit{Docker}\footnote{https://www.docker.com/} containers are used to guarantee that the software testing environment is clean and reproducible. \textit{MATLAB Parallel Server} can be applied to parallelize time-consuming simulations, such as Monte Carlo simulation. Currently, the CI servers are running on discarded desktop computers at the institute. However, one can also choose to run them in the cloud with \textit{Docker} and \textit{Docker Machine}\footnote{https://docs.docker.com/machine/} and gain scalability with minimum additional effort. Fig.~\ref{fig:CI_overview} depicts the overview of the CI process. Based on the contents of a new commit from the developers to the source control server, certain series of tests, which are mostly in the form of \textit{Mrails jobs}, are triggered. The test results are automatically fed back to the developers via \textit{GitLab} and to \textit{Polarion}  via the pre-configured links between the requirements, \textit{Design Models} and the test cases that are described in section \ref{sec:requirementsManagement}. A \textit{git} branch merge into the release or master branch will be rejected if the tests do not pass. This guarantees the quality of the software and allows merges to be performed more often without waiting for manual review.

\subsection{Hardware-in-the-Loop Tests}\label{sec:HIL}
HIL simulations operate a unit-under-test within a representation of its real-world environment \cite{bacicHIL}. Therefore, a target which executes the developed software is connected to a real-time simulation of the later environment the target will be operating in. The basic concept of HIL uses the target's original interfaces for input stimulation from the simulation environment as well as for collecting the target's output into the simulation environment. Thus, a HIL simulation allows to evaluate the target's integrity under real world conditions like input-output interaction, real-time behavior, interface delays and jitters.

Certain objectives in DO-178C pertaining to the verification of Executable Object Code cannot be satisfied with simulation alone and need to be executed on the target environment\cite{vectorDO}. A classical approach is the definition and creation of HIL test cases for requirements-based testing. These test cases are run in HIL simulation and demonstrate, if passed, that the software satisfies the requirements and can be considered as successfully verified \cite{RAY2011277}. 

Since HIL testing is a time-consuming process \cite{RAY2011277}, already available test cases from previous verification methods are reused to reduce the implementation effort.

As already introduced in section \ref{sec:requirementsManagement}, test cases are created in \textit{Simulink Test} and linked to the corresponding requirement. These test cases are used to verify the functional software requirements implemented by the \textit{Design Model} in \textit{Simulink} via MIL simulations. To avoid additional effort in creating dedicated test cases for HIL simulation, thus also reducing an error prone manual transfer between MIL and HIL simulation test cases, \textit{Simulink Test Manger} is used to execute identical MIL test cases in HIL. Besides streamlining the verification process by reducing manual effort, another benefit is that traceability between requirements and HIL test cases is automatically ensured as the test case itself is already linked to the requirement. 
 
\begin{figure}[hbt!]
	\centering
	\includegraphics[width=\columnwidth]{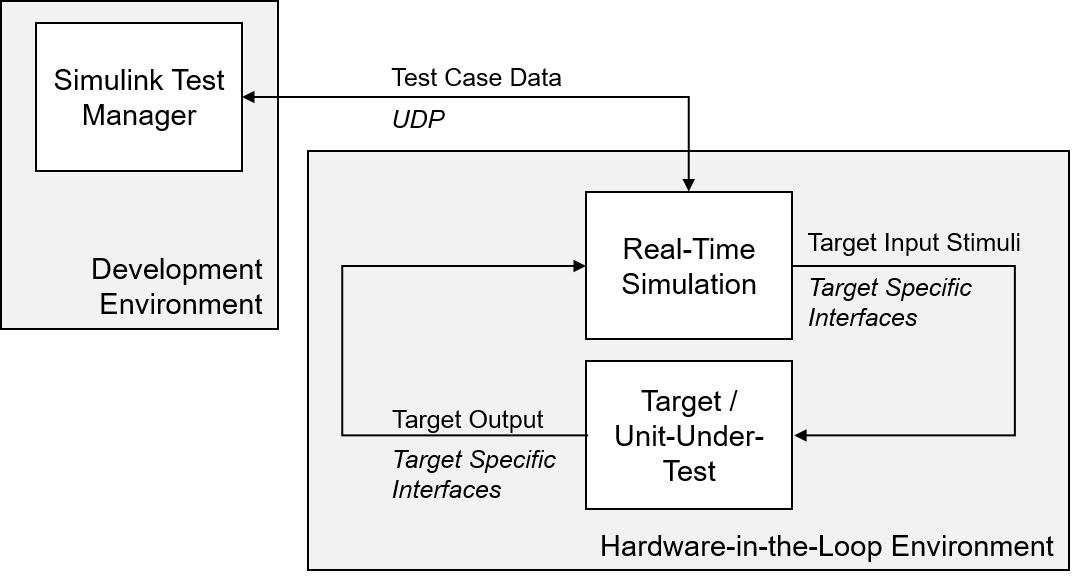}
	\caption{HIL Environment}
	\label{fig:hilEnv}
\end{figure}

In order to be able to apply the same test case as MIL in HIL simulation, the connection of test case vectors to the \textit{Design Model's} inputs and outputs, is substituted by a hardware interface between the development environment which is running the \textit{Simulink Test Manager} and the HIL real-time simulation as shown in Fig.~\ref{fig:hilEnv}.

We have used UDP communication between the \textit{Simulink Test Manager} which is running the test case’s test sequence and the HIL real-time simulation. UDP was selected as a standardized interface between physical machines which can be easily adapted to test cases depending on their input and output vectors also known as test data. \textit{Simulink} furthermore directly provides a block-set for UDP communication. In our example, it is assumed that both the development environment and the HIL real-time simulation are on different computing units or physical machines. The software running on the target platform is receiving the same stimuli from the HIL real-time simulation as the design model receives during MIL simulation. The same holds for its outputs. Using this method, the MIL and HIL test case results are obtained from an identical test case and the results of both tests are supposed to be equivalent as long as the software is implemented correctly.

\section{Level of Compliance}\label{sec:lvlComp}
DO-178C and DO-331 specify various activities, which should be implemented to achieve the objectives of these standards. The artifacts to be produced during the activities is specified in section 11 of DO-178C and DO-331.  Some of the required activities and elements of the artifacts have been considered as not applicable to the \textit{custom workflow}. While implementing the \textit{custom workflow}, we identified and assessed such inapplicable activities and artifact elements. If they were considered as reproducible without major rework of preexisting artifacts, we excluded them from the custom workflow.  Examples are:
\begin{itemize}
	\item Elements of the planning data intended for a formal certification process with certification authorities do not apply to the \textit{custom workflow} (e.g. certification considerations and schedule, system and software description).
	\item Activities of the configuration management process specific to the operation phase of the aircraft (e.g. software part numbering, load control and aircraft configuration records).
\end{itemize}
We also omitted some activities, which cannot be efficiently automated with the tool, however, can be reproduced without major rework afterwards for formal certification. For example:
\begin{itemize}
    \item Completeness of traceability between requirements, models, test cases and results was verified in the \textit{custom workflow} using automated trace reports. However, manual verification of trace data for consistency and correctness was skipped as a time-consuming manual task, which can be implemented afterwards for formal certification.
    \item Verification activities for LLS requirements which cannot be automated and need extensive manual review.
\end{itemize}
Omitted activities or artifact elements resulted in partial compliance for some of the objectives implemented in the \textit{custom workflow}. We identified the following levels of compliance for the selected objectives:
\begin{itemize}
	\item Full – when all activities associated with the objective implemented and artifacts produced
	\item Partial – when some activities associated with the objective are not fully implemented or artifact elements not produced. 
\end{itemize}
A summary of the compliance level achieved in the custom workflow is shown in Fig.~\ref{fig:do331complianceSummary} for DO-331 and Fig.~\ref{fig:do178complianceSummary} for DO-178C.

\section{A Case Study}\label{sec:caseStudy}
The \textit{custom workflow} explained above was applied to the development of  an experimental \textit{Autopilot Manual Disconnection System} (AMDS). The simplicity of the system ensured that a complete workflow from aircraft requirement authoring in Polarion down to HIL testing could be demonstrated with limited project resources.
\begin{figure}[!hbt]
 	\centering
 	\includegraphics[width=\columnwidth]{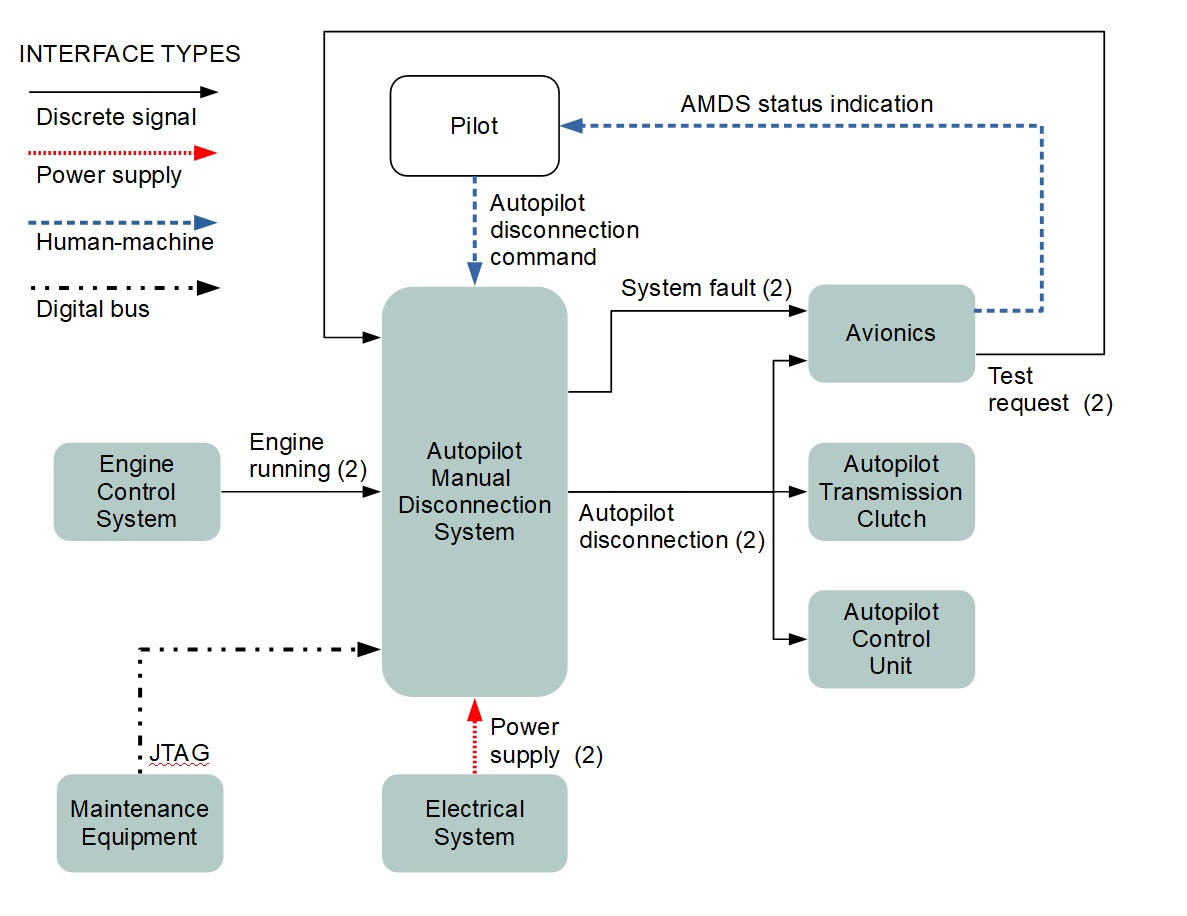}
 	\caption{External Interfaces of AMDS}
 	\label{fig:ADS_interfaces}
 \end{figure} 

Using AMDS, the pilot has aircraft command priority over the autopilot by having the ability to manually disconnect the autopilot at any time during the flight. External interfaces of AMDS are shown in Fig. \ref{fig:ADS_interfaces}. The pilots interact with the AMDS using two pushbuttons, one on each of the control inceptor. Under normal mode of operation, if any of the pushbutton is pressed, the system disconnects the autopilot within 100~ms by controlling the transmission clutch.

A control unit with two isolated and independent channels simultaneously processing the desired function is used. This redundancy ensures that a single failure doesn't lead to loss of primary function. The control unit also implements features typical for airborne controllers, including comprehensive built-in tests and operation-safe maintenance capabilities.

We developed the AMDS software according to the \textit{custom workflow} and tools discussed in section \ref{sec:wrkOvr}. After eliciting the requirements on \textit{Polarion} as described in section \ref{sec:requirementsManagement}, the AMDS functionality is implemented in \textit{Simulink}, whereas test cases assessing the temporal behavior and mode logics based on requirements are authored in \textit{Simulink Test}. Trace data is created between \textit{Simulink} models and \textit{Simulink} test cases, and requirements on \textit{Polarion} using \textit{SimPol}. 

\begin{figure*}[!hbt]
	\centering
	\includegraphics[scale=0.6]{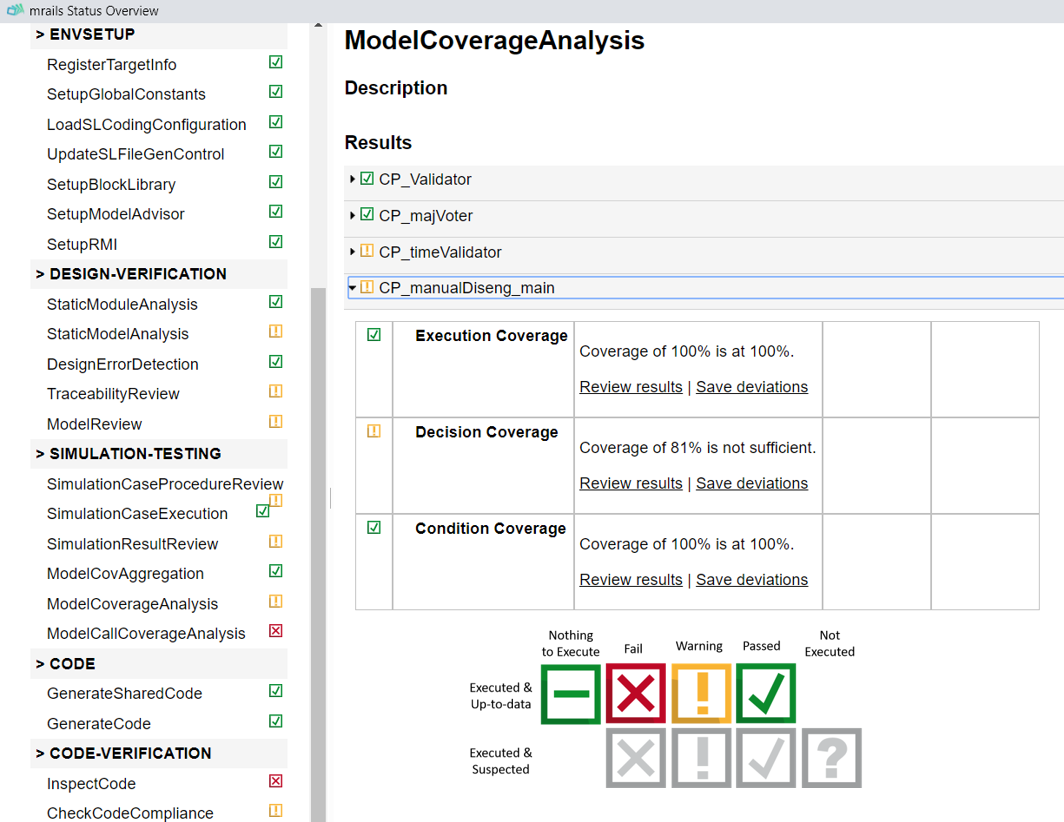}
	\caption{Mrails Status Window}
	\label{fig:mrails}
\end{figure*}
During the development, we used \textit{mrails} described in section \ref{sec:mrails} to execute code generation and all the model and code verification activities shown in Fig. \ref{fig:WF} (excluding HIL testing which is implemented in a real-time environment). A snippet of \textit{mrails}' status overview showing model coverage is given in Fig.~\ref{fig:mrails}. After pushing to \textit{Gitlab}, the CI executes the whole set of tests and static checks to make sure that the modifications are ready to be merged. 

The auto-generated software (MBS) is integrated with LLS and hardware which are reused from the real flight control computer (FCC) described in \cite{nurnberger2017execution}. Building the final loadable executable object code including both MBS and LLS parts is implemented in CI.

To satisfy target specific objectives of requirement-based testing, HIL simulations are performed using TechSAT ADS2\footnote{https://techsat.com/technology-and-products/ads2/}. As described in section \ref{sec:HIL}, MIL test cases are reused for HIL simulations using the UDP interface between HIL and Simulink environment. Equivalence between MIL, SIL and HIL cases have shown that the implemented software works as expected in the target environment and complies with requirements.

By implementing the proposed workflow for the AMDS case study, the following properties of the \textit{custom workflow} were demonstrated:

\begin{itemize}
    \item The overall development effort is estimated to be about 50\% less than typical effort for level C software development by a team without certification experience. Per \cite{thomas2009certCost}, the metric observed in the industry for level C is about 15 lines of code per person per day which includes verification, documentation and other efforts associated with certification. In the case study, the estimated overall performance rate was about 30 lines of code per person per day for the MBS component.
    
    \item High software quality with reduced manual effort. By using early comprehensive testing at model level via simulation with model coverage, no bugs in the MBS component were detected during further SIL and HIL testing. By using modelling guidelines and design error detection earlier for models, no source code defects and only 21 MISRA advisory warnings were found in the MBS code without actively resolving any errors or warnings. 
\end{itemize}

\section{Future Work}\label{sec:futureWrk}
In future work we are planning on upgrading the developed prototype to DO-178C/DO-331 level A and estimate the efforts required to close the gap between the custom workflow and formal DO-178C/DO-331 certification. Another work in progress is applying the custom workflow to a more complex control system to study the compatibility of the approach with a full-scale development project.

For requirements management, we will evaluate the possibility to integrate \textit{FRET} into \textit{Polarion} as an extension. We will also apply the analysis capabilities of \textit{FRET} for linear temporal logics. The custom widget of the automatic trace reports can also be used for coverage reports of model-based validation which will be part of our extended \textit{custom workflow}. The uploaded test run records on \textit{Polarion} are also not yet used for automatic verification reports but this will be implemented in the near future. It is also possible to configure \textit{Polarion} to link manually written \textit{ C/Java/XML } code to work items. These links have not yet been included in the trace reports but they are planned for implementation in the future.

For CI integration, we plan to include automatic HIL tests on the HIL testbed with target hardware in the CI.

\section{Conclusions}\label{sec:conclusions}
The paper has introduced a lean, highly-automated and scalable software development workflow intended for prototype and experimental airborne systems. The proposed workflow is based on the subset of DO-178C and DO-331 objectives which we selected based on criteria of importance, available automation and reuse potential to streamline the development. Extensive automation of development and verification activities using the qualified model-based commercial toolchain along with the own custom tools and continuous integration environment was leveraged to reduce manual development effort while maintaining a high software quality. This \textit{custom workflow} is aimed to be used a basis for potential future certification in full compliance with DO-178C/DO-331 without wide rework of the produced data.

Using the proposed \textit{custom workflow}, we developed a small system prototype to study the advantages and limitations of this method. We showed that development fulfilling the selected DO-178C objectives using the \textit{custom workflow} can reduce development efforts by about 50\% comparing to average metrics observed in the industry for DO-178C level C software. Yet, a high software quality was achieved with respect to the number of bugs detected and compliance to coding guidelines. 
	
We plan in our future work to estimate the efforts needed to close the gap between the \textit{custom workflow} and level A software development and also explore the relevance of the workflow to a full-scale development of a complex system.

\bibliographystyle{IEEEtran}
\bibliography{references}

\end{document}